# Phonon-modulated Kerr nonlinearity in ultrathin 2$H$-MoTe$_2$

Shaoxiang Sheng[1,2]†*, Yang Luo[1,3]†, Chenyu Wang[4,5]†, Sayooj Sateesh[1], Yaxian Wang[4], Marko Burkhard[1], Sayantan Patra[6], Bhumika Chauhan[6], Ashish Arora[6], Sheng Meng[4,5]*, Manish Garg[1,*]

[1]Max Planck Institute for Solid State Research, Heisenbergstr. 1, 70569 Stuttgart, Germany

[2]Tsientang Institute for Advanced Study, Hangzhou, 310024, China

[3]Hefei National Laboratory, University of Science and Technology of China, Hefei, 230088, China

[4]Beijing National Laboratory for Condensed Matter Physics and Institute of Physics, Chinese Academy of Sciences, Beijing 100190, China

[5]School of Physical Sciences, University of Chinese Academy of Sciences, Beijing 100190, China

[6]Indian Institute of Science Education and Research, Pune, 411008, India

*Corresponding authors. Email: s.sheng@fkf.mpg.de, smeng@iphy.ac.cn, mgarg@fkf.mpg.de

†These authors contributed equally to this work: S. Sheng, Y. Luo, and C. Wang

**Controlling nonequilibrium responses in optically driven quantum materials is essential for advancing applications in energy conversion, ultrafast electronics, and quantum computation. Nonlinear optical spectroscopy serves as a powerful tool to investigate ultrafast electron and phonon dynamics in these systems; however, conventional nonlinear approaches often require intense laser pulses (> 10 GW/cm$^2$) and typically encounter a strong background. Here, we introduce a phase-sensitive nonlinear spectroscopic technique that operates at low laser powers (~ 10 kW/cm$^2$, pulse energies ~ 10 pJ) and enables real-time monitoring and active control of coherent phonons in a few-layer (three to five) thick 2$H$-MoTe$_2$. Upon excitation with ultrashort (~ 10 fs) pump pulses, we achieve displacive excitation of coherent phonons, which periodically modulate the Kerr nonlinearity of the material, leading to cross-phase modulation (XPM) of a delayed probe pulse. This phase modulation induces spectral broadening and oscillations in the center of mass (COM) of the probe spectrum in time, enabling the detection of subtle nonlinear optical responses in a background-free manner. The nonlinear response can be selectively amplified or attenuated by adjusting the strength of the pump pulse, which controls the distribution of photoexcited carriers in the electronic bands. By combining two-color nondegenerate pump-probe measurements and time-dependent density-functional theory (TDDFT) calculations, we directly resolve the coupled nonequilibrium electronic and phonon dynamics. A dual-pump pulse scheme enables precise control of phonon oscillations, allowing selective activation or suppression of specific phonon modes and correspondingly the modulation of the Kerr nonlinearity. Our work provides new perspectives of manipulating quantum materials on femtosecond timescales, particularly in phonon-mediated phenomena, such as transient superconductivity and structural phase transitions.**

Ultrafast optical control of quantum materials provides a powerful approach to access nonequilibrium states of matter[1,2], such as light-induced superconductivity[3,4], magnetism[5], and topological phases[6,7]. Understanding these transient phenomena requires insights into the complex interplay between electronic excitations and coherent phonon motions, yet, state-of-the-art techniques struggle to disentangle these dynamics at their natural timescales while providing phase-sensitive information.

Nonlinear optical processes, e.g., the Kerr effect—a third-order nonlinear ($\chi^{(3)}$) response leading to intensity-dependent modulation of the refractive index—are crucial in ultrafast signal processing and in the investigation of complex light-matter interactions[8–11]. In particular, cross-phase modulation (XPM), where a 'pump' pulse-induced nonlinearity modulates the phase of a time-delayed 'probe' pulse via the nonlinear Kerr effect, offers a highly phase-sensitive approach for tracking both coherent and incoherent quasiparticle dynamics[12]. In contrast to self-phase modulation (SPM), which entangles various nonlinear interactions within the material[13], XPM provides a phase-locked detection tool of dynamics evolving in real-time. In this process, the intensity, polarization, and the frequency of the probe pulse can be dynamically modulated via interactions with the material pre-excited by an ultrafast pump pulse[14–16].

Leveraging XPM for ultrafast material control is particularly compelling in layered transition metal dichalcogenides (TMDCs), which exhibit pronounced Kerr nonlinearities alongside exceptional optical and electronic properties[17–19]. Among them, MoTe$_2$ is distinguished by its rich nonequilibrium states and structural phase transitions under intense optical pulse excitation[20–23], making it a promising platform for all-optical modulation and control.

**Phonon-mediated cross-phase modulation**

In this work, we introduce a phase-sensitive XPM spectroscopic technique that directly resolves and controls femtosecond electron and phonon dynamics in a few-layer thick 2H-MoTe₂. Ultrashort laser pulses generated by an ultra-broadband Ti:Sapphire oscillator were split into two spectrally separated 'pump' and 'probe' pulses using an 800 nm shortpass dichroic mirror (DMSP), Fig. 1a. The pump (λ ~800–950 nm, ~10 fs long) and probe pulses (λ ~700–750 nm, ~20 fs long) were further spectrally filtered using bandpass (BP) filters to avoid any spectral overlap and interference artifacts between them, their respective spectra are shown in Fig. 1b. The probe pulse was delayed using a precision motorized linear stage before being recombined with the pump pulse. The collinearly propagating pump and probe pulses were then focused onto the sample using a reflective objective (numerical aperture ~ 0.5). The light emanating from the sample was collected using the same objective and passed through a 690 nm shortpass filter (SP) before being guided into a spectrometer.

A probe pulse propagating through the 2H-MoTe₂ sample experiences a time-varying refractive index[24] owing to the Kerr effect, thus, inducing a time-dependent phase shift across the pulse envelope[25]:

$$\Delta\phi(t) = \frac{2\pi}{\lambda} n_2 I(t) L \qquad 1)$$

where $\lambda$ is the wavelength of incident light, $n_2$ is the nonlinear refractive index coefficient of 2H-MoTe₂, $I(t)$ is the instantaneous intensity of the probe pulse, and $L$ is the effective interaction length within the material. This phase shift leads to spectral broadening (see Fig. S1), characterized by the instantaneous frequency change:

$$\delta\omega(t) = -\frac{\partial \Delta\phi(t)}{\partial t} \qquad 2)$$

When both the pump and probe pulses co-propagate through the nonlinear medium, the phase shift experienced by the probe pulse is influenced by the combined intensity of both pulses. Consequently, the time-dependent phase shift acquired by the probe pulse becomes:

$$\Delta\phi(t) = \frac{2\pi}{\lambda} n_2 \left( I_{probe}(t) + \kappa I_{pump}(t) \right) L \qquad 3)$$

where $\kappa$ is a coefficient, often close to 2 in different materials[25–27], indicating that the pump pulse exerts a stronger influence on the probe pulse, than the probe pulse does on itself (SPM). This interaction gives rise to an intense XPM signal at the zero time delay between the pulses, as shown by the green-shaded spectral region in Fig. 1b and the inset (right) sketch. Moreover, when the pump and probe pulses temporally overlap, four-wave mixing (FWM) processes also occur. One prominent case involving two photons from the probe pulse and one photon from the pump pulse results in the blue-shaded spectrum in Fig. 1b, as depicted in the inset (left).

The nonlinear XPM and FWM responses were characterized by varying the delay between the pump and the probe pulses. Figure 2a shows a series of pump-probe spectra, in which the individual spectra were subtracted with a spectrum measured at a far negative delay of -1 ps between the pulses, effectively removing the SPM signal generated by the probe pulse alone. Owing to the broad spectral separation between the pump and probe pulses, SPM induced solely by the pump pulse is not detected. At zero delay, pronounced nonlinear signals emerge, corresponding to XPM (~ 670–690 nm) and FWM (~ 620–670 nm). Following the initial pump-probe overlap, the XPM signal exhibits a sharp drop in its amplitude followed by periodic oscillations lasting up to ~ 10 ps. The initial decrease and recovery of the XPM signal at longer delays is attributed to an ultrafast electronic response, where depletion of the carriers in the valence band upon photoexcitation with the pump pulses reduces the Kerr nonlinearity[28].

To analyze the oscillatory behavior in detail, the temporal crosscut of the pump-probe XPM spectra at ~ 680 nm is plotted in Fig. 2b. The time-trace exhibits a strong peak at the zero delay, followed by an abrupt reduction and a subsequent recovery within ~1 ps, superimposed with long-lived oscillations. Fast Fourier transform (FFT) analysis of this time-domain signal reveals a pronounced peak at ~ 172 cm$^{-1}$ (upper panel in Fig. 2c), corresponding to the out-of-plane $A_{1g}$ phonon mode of 2$H$-MoTe$_2$. This mode exhibits a long coherence lifetime (~ 5.0 ps), consistent with previous transient reflection measurements[29]. In addition, our data also reveals a broad peak centered at ~ 480 cm$^{-1}$, indicating a short-lived phonon mode (~ 0.4 ps, see also the time-frequency analysis of the pump-probe traces in Fig. S2 in SI). This spectral peak is attributed to the $A_{1b}$ mode of the quartz substrate, a ring O$_{bridge}$-breathing phonon mode[30].

Apart from these dominant phonon modes, we also observed a weak spectral peak at ~ 202 cm$^{-1}$ (Fig. 2c) that matches the $A_{1r}$ (rotation mode of SiO$_4$ tetrahedra) mode of quartz[30]. These mode attributions were further confirmed by Raman spectroscopic characterization performed on the same sample (lower-panel in Fig. 2c). Notably, few-layer 2$H$-MoTe$_2$ transferred onto a Si$_3$N$_4$ substrate shows neither the $A_{1b}$ nor the $A_{1r}$ mode (see Fig. S3 in SI). The presence of the phonon modes of the quartz substrate in the time-resolved measurements corroborates the exceptional phase sensitivity of the present XPM-based technique.

A cosine-like oscillation can be assigned to both the $A_{1g}$ mode of 2$H$-MoTe$_2$ and $A_{1b}$ mode of quartz by fitting the time-domain signal backwards to the zero delay with phases of $(0.02 \pm 0.07)\pi$, and $(1.02 \pm 0.05)\pi$, respectively, as shown in the inset of Fig. 2b. The absolute maximum of the FWM signal in the measurements enables clocking of the zero time-delay between the pump and the probe pulses. The opposite phases at the zero-delay of the two phonon modes likely results from their distinct eigenvector orientations and differences in the underlying band structures[30,31]. Notably, the strongest Raman-active $E_{2g}$ mode of 2$H$-MoTe$_2$ at ~240 cm$^{-1}$ is absent in the time-resolved XPM measurements (Fig. 2c). A similar behavior was observed in the few-layer thick 1$T$-MoTe$_2$, where the $A_g$ mode at ~ 253 cm$^{-1}$ is observed without the presence of the corresponding $E_{2g}$ mode (see Fig. S3 in SI). Coherent phonon oscillations corresponding to the $A_g$ mode were also measured for a monolayer 2$H$-MoTe$_2$ mechanically exfoliated onto a z-cut sapphire substrate (see Fig. S4 in SI). The selective mode excitation of fully symmetric $A$ modes and their cosine-like oscillatory behavior suggest a displacive excitation mechanism of phonons in MoTe$_2$ and the underlying quartz substrate[32], as schematically shown in Fig 2e. Illumination with ultrafast laser pulses induces an instantaneous redistribution of electronic charge in 2$H$-MoTe$_2$. This rapid electronic redistribution alters the potential energy surface (PES), $U$, on attosecond timescales[30], causing a shift in the equilibrium position ($Q_0$) of the lattice within the adiabatic Born–Oppenheimer approximation. As a result, atoms are impulsively displaced from their original equilibrium positions and subsequently oscillate coherently around the new equilibrium position ($Q_0$') defined by the transient electronic state[31].

The oscillations in the XPM signal at the phonon frequencies of 2$H$-MoTe$_2$ and quartz indicates that the pump pulses excite coherent phonons, which are subsequently detected by the probe-pulse via phase-sensitive nonlinear interactions. This interpretation is further attested by time-dependent density functional theory (TDDFT) calculations, see Fig. S5 and SI for details. The probe pulse on interacting with the coherently oscillating lattice acquires a periodic phase modulation, leading to a periodic shift in the COM of its spectral weight. This spectral shift manifests experimentally as oscillations in the broadened spectral intensity of the probe-pulse[12]. By incorporating both FWM and XPM contributions into the simulations, the key features in the pump-probe spectra can be reasonably reproduced, as shown in Fig. 2d. The transient modulation of the Kerr nonlinearity by the pump pulse can be attributed to photoexcitation of the carriers and the resulting modifications of the electronic band structure[9,28], which will be discussed later in the text. In the simulation shown in Fig. 2d, the ultrafast electronic depletion is modelled using a step-like response followed by an exponential decay[9,28,32], see methods for details.

Presence of extensive dynamical features in the pump-probe traces demonstrate that the probe pulse is highly sensitive to pump-induced modifications of the dielectric function of the material[33]. Compared to other nonlinear techniques, such as high-harmonic generation[30,34], transient reflection and transmission spectroscopy[35], this phase-sensitive approach operates at substantially lower pulse energies (< 10 pJ) while providing superior phase sensitivity. As a result, it offers a powerful tool for resolving quasiparticle interactions, exciton formation, and their relaxation dynamics in quantum materials.

**Displacive excitation of symmetric phonon modes**

To elucidate the mechanism underlying the coherent phonon excitation, we performed fluence-dependent measurements. First, we varied the fluence of the probe pulses while keeping the pump fluence fixed. The spectral intensities of the retrieved phonon modes, $A_{1g}$ and $A_{1b}$, exhibit a quadratic dependence on the probe fluence, as shown in Fig. 3a. Similarly, the amplitude of the depleted XPM signal at the zero pump-probe delay, also exhibits a quadratic dependence on the probe fluence (blue curve, Fig. 3b). The retrieved quadratic scaling of the intensity of the phonon modes rules out the FWM (or coherent anti-Stokes Raman (CARS)) process involving two photons from the pump pulse and one photon from the probe pulse as the origin of the vibrational coherence (see also Fig. S6)[36,37]. It is worth mentioning that the anti-Stokes Raman signal would be fully suppressed in our experimental configuration (see SI for details). The observed FWM process at ~ 650 nm, involving two photons from the probe pulse and one photon from the pump pulse, exhibits the expected quadratic scaling on probe fluence (see Fig. S7). The nonlinear dependence of both the XPM response and phonon amplitudes on probe fluence strongly supports the nonlinear XPM mechanism. It is worth noting that the decay constants associated with the relaxation times of the photocarriers exhibit a decreasing trend on increasing probe fluence (red curve in Fig. 3b), reflecting the role of enhanced carrier-carrier interactions at higher photoexcitation intensities on the retrieved relaxation times.

The excitation of coherent phonons on interaction with ultrashort laser pulses typically arises from two plausible mechanisms: (1) impulsive stimulated Raman scattering (ISRS) or (2) displacive excitation of coherent phonons (DECP)[29]. To further distinguish between them, we performed time-resolved XPM measurements by varying the pump fluence, while keeping the fluence of the probe pulse fixed. The temporal crosscuts at ~ 680 nm from the pump-probe spectra are shown in Fig. 3c. Compared to the probe fluence dependence measurement, the time-resolved traces show markedly more complex behavior (see time traces in Fig. S8). At low fluence of the pump pulse, the XPM signal gradually increases after the zero time delay (bottom curves in Fig. 3c), in contrast with the abrupt reduction observed in Fig. 2a and Fig. 2b.

As the pump fluence increases, this gradual rise in the XPM signal, after the zero time delay, transitions into a pronounced depletion (top curves in Fig. 3c). This amplification or attenuation of the XPM signal directly relates to an enhanced or reduced optical Kerr nonlinearity ($n_2$) on varying the pump fluence[38]. While this behavior complicates the extraction of the decay constants of the electronic response, phonon amplitudes can still be robustly quantified via the FFT analysis of the time-domain traces. The extracted amplitudes of both the $A_{1g}$ and $A_{1b}$ modes exhibit virtually a sub-linear dependence on the pump fluence (Fig. 3d), which further substantiates the DECP mechanism[32].

Within the DECP framework, the uniform photoexcitation of electrons creates a symmetric driving force: $F(t) = -\frac{\partial U(t)}{\partial Q}$, leading to the selective excitation of fully symmetric vibrational modes, such as the $A_{1g}$ mode in MoTe$_2$, whose atomic displacements match the symmetry of the perturbation. In contrast, low-symmetry modes, such as doubly degenerate $E$-type modes, which involve antisymmetric or degenerate atomic motions, are not efficiently excited, as their displacements do not align with the symmetry of the driving force[39]. This symmetry-selective excitation mechanism explains the exclusive observation of fully symmetric $A$ modes in both 2$H$-MoTe$_2$ and quartz (a similar trend is also observed in 1$T$-MoTe$_2$, see SI for details).

**Amplifying and attenuating Kerr nonlinear response**

We now focus on the amplification and attenuation behavior of the Kerr nonlinear response induced by the nonlinear interaction of the probe pulse ensuing excitation by the pump pulse, as highlighted by the red and blue areas in Fig. 3c. The interband contribution to the imaginary part of the third-order nonlinear susceptibility, $\chi^{(3)}_{inter}$, can be expressed in terms of the electronic distribution in different bands as[40]:

$$Im\chi^{(3)}_{inter} \propto \sum_{i,j} \int \frac{d^3k}{4\pi^3} |P_{ij}(\boldsymbol{k})|^4 \times f_i(\boldsymbol{k})\left(1 - f_j(\boldsymbol{k})\right) \times \delta\left(E_j(\boldsymbol{k}) - E_i(\boldsymbol{k}) - \hbar\omega\right) \qquad 4)$$

where $P_{ij}$ denotes the matrix element of the momentum operator, and $f_j$ and $f_j$ are the occupation probabilities of the initial ($i$) state and the final ($j$) states. Thus, photoexcitation directly modifies the nonlinear refractive index $n_2$ via changes in the electronic population. In the present case, we focus on the $K_1$ and $K_2$ points in the band structure, where the pump pulse (~ 1.3 – 1.55 eV) resonantly excites electrons from the valence band to the conduction band, as indicated by the vertical red arrows in Fig. 3e.

In this case, the pump-induced modification in Kerr nonlinearity mainly arises from electronic populations at the $K_1$ and $K_2$ points. At the $K_1$ point, the valence band states deplete under resonant pump excitation (blue area on the left side of Fig. 3e), nonetheless, the remaining electrons at the $K_1$ point can still participate in XPM process via resonant interband transitions (solid green arrow on the left side of Fig. 3e). However, this transition probability would be reduced on increasing the pump fluence, because a fraction of valence-band electrons would have already been promoted to the conduction band. This results into the reduction in the $\chi^{(3)}$ contribution and leads to a suppression of the nonlinear signal, consistent with the reduced XPM response observed in Fig. 2a and Fig. 2b.

Nevertheless, photoexcited electrons in the conduction band, unoccupied in equilibrium, can amplify the nonlinear response. This amplification occurs predominantly at the $K_2$ point, where pump-excited carriers (red area on the right side of Fig. 3e) enable resonant probe-induced $\chi^{(3)}$ processes involving higher conduction band states (solid green arrow on the right side of Fig. 3e). This contribution effectively increases the Kerr nonlinearity and the associated XPM signal.

The interplay between the above two processes gives rise to the non-monotonic fluence dependence recorded in Fig. 3c. Upon weak pump pulse excitation, photoexcited carriers at the $K_2$ point ($n_{K2}$) enhance the nonlinear response experienced by the probe pulse, whereas the carrier population in the conduction band near $K_1$, remains largely unchanged. On increasing the pump fluence, resonant depletion of valence band states near $K_1$ point ($n_{K1}$) dominates, leading to a net reduction of the XPM signal. This imbalance in the carrier population, $\Delta n = n_{K2} - n_{K1}$, as captured by the TDDFT calculations, which increases at low pump fluence (red area) and decreases at higher fluence (blue area), is shown by the black squares in Fig. 3f.

In contrast to the complex nonlinear electronic response, we find that the phonon amplitude increases virtually linearly with pump fluence, as shown by the red dots in Fig. 3f, in agreement with the experimentally extracted phonon amplitudes shown in Fig. 3d. These results suggest that electronic and lattice dynamics can be independently and sensitively controlled using even weak optical gating pulse[30,32,39]. Such low-power control of optical nonlinearity is highly desirable for ultrafast modulation, signal processing and photonic switching applications.

**Coherent control of phonon oscillations via a dual pump scheme**

Having established the excitation and detection mechanisms of the coherent phonons, we now focus on their controlled manipulation, which is crucial in shaping the physical properties of materials on the terahertz timescales. Controlled phonon dynamics is particularly relevant in phenomena such as charge density wave formation and electron pairing in superconductors. To achieve this, we employed two identical pump pulses, hereafter referred to as 'pump-1' and 'pump-2', each capable of individually exciting coherent phonons. By precisely tuning the delay between these two pump pulses, phonon oscillations can be selectively enhanced or suppressed through constructive or destructive interference of the launched phonon wavepackets, as illustrated in Fig. 4a. The phase-sensitive modulation of the probe pulse enables real-time monitoring of this interference of the phonon wavepackets.

Figure 4b demonstrates phonon control by varying the delay, $T$, between the pump-1 and pump-2 pulses. In each time trace, the delay between these two pump pulses were fixed as labelled along the vertical y-axis, whereas the delay between the two pump pulses and the probe pulse was varied along the horizontal x-axis. The individual time traces in Fig. 4b represent temporal crosscuts at ~ 680 nm from the pump-probe measurements. The amplitudes of the phonon oscillations exhibit a clear periodic modulation as a function of $T$. The FFT spectra (Fig. 4c) of these time traces reveal that the amplitude of the $A_{1g}$ mode oscillates with a period of $T_{ph}$ ~ 194 fs, corresponding to the $A_{1g}$ phonon mode frequency (~ 172 cm$^{-1}$). At specific pump-pump delays, the oscillations are largely enhanced ($T \sim 2T_{ph}$, red arrows in Fig. 4b), whereas at other delays, they are almost completely suppressed ($T \sim 2.5T_{ph}$, purple arrows in Fig. 4b). The representative time traces and their FFT spectra (inset) are shown in Fig. 4d.

In addition to the amplitude control, the phase of the phonon oscillations can also be tuned by adjusting the delay between the pump-1 and pump-2 pulses, allowing the oscillations to be driven completely in- or out-of-phase, as shown in Fig. 4e. This behavior is extracted from the time traces indicated by the orange and yellow arrows in Fig. 4b. The interference of the phonon wavepackets leading to variation in the amplitude (red dots) and phase (blue dots) of the $A_{1g}$ phonon mode can be modeled using two driven damped harmonic oscillators[41]:

$$Q_v(t,T) = Q_{v,0} \cos(\omega_v t)\, e^{-\frac{t}{\tau_p}} + Q_{v,0} \cos(\omega_v(t-T))\, e^{-\frac{(t-T)}{\tau_p}} \qquad 5)$$

where $Q_{v,0}$ is the initial amplitude set by DECP excitation, $\omega_v$ is the phonon oscillation frequency, and $\tau_p$ is the phonon decay time constant. The resulting phase of the phonon oscillation is given by:

$$\Phi_v(T) = -\arctan\left(\frac{\sin(\omega_v T)\, e^{\frac{T}{\tau_p}}}{\cos(\omega_v T)\, e^{\frac{T}{\tau_p}} + 1}\right) \qquad 6)$$

Using this model, we accurately reproduce the experimentally observed amplitude and phase modulation, as shown by the solid curves in Fig. 4f. The phonon control as experimentally measured is also independently confirmed by TDDFT calculations (see SI for details).

This phonon control offers a novel approach for manipulating dynamics in materials. By selectively turning phonon oscillations on or off, the electronic response can also be modulated, as observed by the gain in the XPM signal—indicating an enhanced Kerr nonlinearity—observed at shorter delays between the pump-1 and pump-2 pulses in Fig. 4b. The relatively slow modulation of the Kerr nonlinear response leads to the emergence of a very low-frequency mode in the FFT spectra in Fig. 4c. Such a precise control over phonons holds promise for manipulating nonequilibrium phases in MoTe$_2$, such as phonon-mediated superconductivity and phase transitions[22,42].

**Concluding Remarks**

In this work, we demonstrated a phase-sensitive nonlinear optical spectroscopic technique capable of providing direct insights into the ultrafast electronic and phononic dynamics. We showed that the optically excited coherent phonons modulate the Kerr nonlinearity of materials, inducing cross-phase modulation on a co-propagating probe pulse. This approach operates with minimal pulse power, is applicable to various materials, and is highly sensitive to transient phase transitions and excitonic dynamics in a nondestructive manner. Coherent control of phonon oscillations was achieved using a dual-pump pulse scheme, enabling precise real-time manipulation of phonon dynamics. This capability lays the foundation for further exploration of phonon-mediated control in quantum materials, such as charge density waves and superconductivity. Our findings open new avenues for phonon-engineered light-matter interactions on femtosecond timescales, and highlight the potential of ultrafast nonlinear optical modulations in advancing next-generation quantum and optoelectronic devices.

**Methods**

**Experimental setup**

Ultrafast pump (λ ~800–950 nm, ~10 fs long) and probe pulses (λ ~700–750 nm, ~20 fs long) were employed for the phase-sensitive nonlinear spectroscopy measurements. The pump and probe pulses were spectrally separated from the ultra-broadband output of a Ti:Sapphire oscillator (Element 2) using an 800 nm shortpass dichroic mirror (DMSP) and were pre-compressed with a pair of chirped mirrors.

The collinearly propagating pump and probe pulses were focused onto the sample (mounted onto a motorized 3D stage) using a 40 × reflective objective with a numerical aperture of 0.5. The light emanating from the sample was then collected using the same objective and passed through a 690 nm shortpass filter (SP) before being directed into a spectrometer for analysis. A precision motorized linear stage (PI Instruments) was used to adjust the delay between the pump and probe pulses. For all spectra, an acquisition time of 0.5 s per delay step was used unless otherwise specified.

For pump-probe experiments, the power and polarization of both pump and probe pulses could be independently controlled using two motorized neutral density (ND) filters in their respective beam paths. Raman spectroscopy was performed on a commercial Raman spectrometer from WiTec, and a 633 nm CW laser was used for the measurements. The laser power was kept below 1 mW to avoid sample damage. All the experiments were conducted at room temperature.

**Samples preparation**

The few-layer (three to five) thick 2$H$-MoTe$_2$ samples, grown by chemical vapor deposition (CVD) on a quartz substrate, were purchased from 2D Semiconductors. The 2$H$-MoTe$_2$ and 1$T$-MoTe$_2$ flakes (in Fig. S3), were mechanically exfoliated from their bulk crystals (bought from HQ Graphene) and transferred onto the Si$_3$N$_4$ substrate within a glove box. The monolayer 2$H$-MoTe$_2$ were mechanically exfoliated and transferred onto a sapphire substrate, and the thicknesses were characterized using Raman spectroscopy (as shown in Fig. S4).

**Pump-probe spectra simulation**

In the pump-probe experiments, nonlinear cross-phase modulation (XPM) and four-wave mixing (FWM) processes occur as the pump and probe pulses interact in the nonlinear medium. For the XPM spectra simulation in Fig. 2d, the Kerr nonlinearity response was separated into phonon and electronic contributions[32]. The phase response modulated by phonon oscillations was modeled as a damped oscillator:

$$\Phi_{p,v}(t,\tau) = a \cdot \cos(\omega_v(t+\tau) + \varphi_v) e^{-\frac{t+\tau}{\tau_p}}$$

where $a$ is a perfector, $\omega_v$ is the phonon oscillation frequency, $\varphi_v$ is the phase of the phonon oscillation, and $\tau_p$ is the phonon decay time constant determined from the phonon decay analysis in Fig. S2.

The phase response contributed by electronic components was modeled as a simple step-decay exponential function upon the pump pulse excitation:

$$\Phi_e(t,\tau) = b \cdot u(\tau) e^{-\frac{t+\tau}{\tau_e}}$$

where $b$ is a prefactor, $u(\tau)$ ($u(\tau \leq 0) = 0, u(\tau > 0) = 1$) is a step function, and $\tau_e$ is the electronic response decay time constant determined from the exponential fit in Fig. 2b. In the simulation, $a$ and $b$ are the only free fit parameters, which are related to the pump laser power.

FWM involves the interaction of the fields with nonlinear polarization of the medium, which results in the generation of new optical fields. In the simulations, the FWM processes that could generate frequencies larger than 700 nm were considered, as shown in Fig. S6. The nonlinear polarization in the medium responsible for the FWM processes could be expressed in terms of the third-order susceptibility $\chi^{(3)}$, which is given by:

$$P = \sum_{j,k,l} \epsilon_0 \chi^{(3)}_{ijkl} E_j E_k E_l$$

where $i$, $j$, and $k$ represent the pump and probe pulse indices, respectively. The generated field at $\omega_i = \pm\omega_j \pm \omega_k \pm \omega_l$ is determined by the pump and probe field frequencies.

By adjusting the free parameters $a$ and $b$ in the XPM simulation, both the FWM and phonon oscillations in Fig. 2d could be reproduced by the model described above.

**Time-dependent density-functional theory (TDDFT)**

The excited state dynamics can be simulated via a real-time density functional theory framework (rt-TDDFT), which we implemented in the QUANTUM ESPRESSO package[43–45].

Under the theorem of TDDFT, the time evolution of electron wave functions is governed by the time dependent Kohn-Sham (TDKS) equations[46]:

$$i\frac{\partial}{\partial t}\psi_{i,\mathbf{k}}(\mathbf{r},t) = \left[\frac{1}{2m}\left(\mathbf{p} - \frac{e}{c}\mathbf{A}\right)^2 + V(\mathbf{r},t)\right]\psi_{i,\mathbf{k}}(\mathbf{r},t),$$

where velocity gauge is used and the external field appears in the kinetic term in the form of vector potential $\mathbf{A}(t)$. The propagation of TDKS orbitals is implemented on the adiabatic basis $\phi_{n,\mathbf{k}}(\mathbf{r},t)$:

$$|\psi_{i,\mathbf{k}}(\mathbf{r},t)\rangle = \sum_n c_{in\mathbf{k}}(t) |\phi_{n,\mathbf{k}}(\mathbf{r},t)\rangle,$$

where $i$ and $n$ denote the band index, $\mathbf{k}$ refers to the reciprocal momentum index and $c_{in\mathbf{k}}(t)$ the time dependent coefficients. The adiabatic basis is calculated on the fly at each ionic step by diagonalizing the Hamiltonian:

$$H_{\mathbf{k}}(\mathbf{r},t)|\phi_{n,\mathbf{k}}(\mathbf{r},t)\rangle = \varepsilon_{n,\mathbf{k}}(t)|\phi_{n,\mathbf{k}}(\mathbf{r},t)\rangle$$

where $\varepsilon_{n,\mathbf{k}}(t)$ is the eigenvalue. The charge density can be calculated with $c_{in\mathbf{k}}(t)$ and $\phi_{n,\mathbf{k}}(\mathbf{r},t)$:

$$\rho(\mathbf{r},t) = \sum_{\mathbf{k}}\sum_n q_{n,\mathbf{k}}(t)|\langle\phi_{n,\mathbf{k}}(\mathbf{r},t)|\phi_{n,\mathbf{k}}(\mathbf{r},t)\rangle|^2,$$

where

$$q_{n,k}(t) = \sum_i |c_{ink}(t)|^2,$$

is the population on the adiabatic basis.

For ions that are much heavier than electrons, their motions are treated classically on an averaged potential energy surface determined by the electronic distribution according to the Ehrenfest theorem. The nuclear positions are updated following the Newton's second law[47]:

$$M_\alpha \frac{d^2 \mathbf{R}_\alpha}{dt^2} = \sum_i \left\langle \psi_i \middle| \nabla_{\mathbf{R}_\alpha} \left( \frac{1}{2m} \left( \mathbf{p} - \frac{e}{c}\mathbf{A} \right)^2 + V(\mathbf{r},t) \right) \middle| \psi_i \right\rangle$$

where $M_\alpha$ and $\mathbf{R}_\alpha$ are the mass and position of the $\alpha th$ ion.

**Data availability**

The data that supports the findings of this study are available from the corresponding authors on request.

**Code availability**

The details needed to reproduce the computations have been provided in the "Methods" section and Supplementary Information file.

**Acknowledgments**

We thank Wolfgang Stiepany and Marko Memmler for technical support.

**Contributions**

S.S., Y.L., Sa.S., M.B., A.A. and M.G. built the experimental setup, performed the experiments and analyzed the experimental data. C.W., Y.W. and S.M. designed and performed the theoretical calculations and analyzed the theoretical data. M.G. conceived the project and designed the experiments. All authors interpreted the results and contributed to the preparation of the manuscript.

**Competing interests**

The authors declare no competing interests.

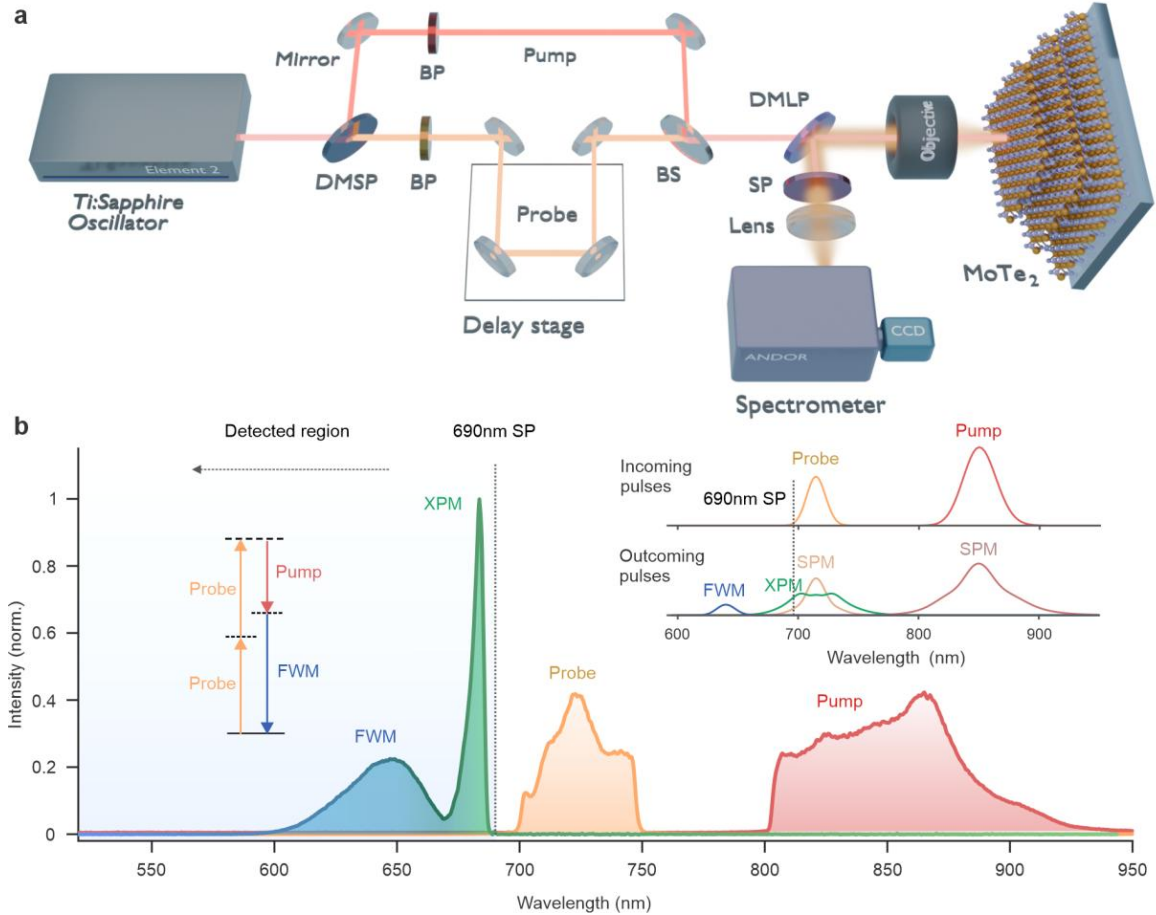

**Figure 1 | Phase-sensitive nonlinear spectroscopy of ultrathin 2*H*-MoTe$_2$. a**, Schematic illustration of phase-sensitive nonlinear pump-probe spectroscopy in a chemical vapor deposition (CVD) grown few-layer (three to five) thick 2*H*-MoTe$_2$ on a quartz substrate. Collinearly propagating nondegenerate pump (~ 800-950 nm, ~ 10 fs long) and probe (~ 700-750 nm, ~ 20 fs long) laser pulses were focused onto the sample using a reflective objective. The light emanating from the sample was collected using the same objective, filtered through a ~ 690 nm shortpass filter to remove residual incident laser components, before being focused into the spectrometer. BP: Bandpass filter, SP: Shortpass filter, DMSP: Shortpass dichroic mirror, DMLP: Longpass dichroic mirror. **b**, Spectra of the incoming pump (red filed curve) and the probe (orange filed curve) pulses, shown on the right side of the dashed black curve, together with the measured spectral broadening of the probe pulse induced by cross-phase modulation (XPM, green filled curve) and the four-wave mixing (FWM, blue filled curved) processes on the left side of the dashed black curve. Left inset: schematic of the FWM contribution to the measured spectra, resulting from interaction of two photons from the probe pulse (upward transitions) ensued by one photon interaction from the pump pulse (downward transition). Right inset: Spectral-domain illustration of the self-phase modulation (SPM), XPM, and the FWM processes. Incoming pump and probe pulses undergo SPM as well as XPM, resulting in broadened spectra. Spectral components present on the left side of the dashed black curve were measured in the experiments, comprising of SPM of the probe pulse, XPM, and the FWM signals. SPM of the pump pulse was not measured as it lies outside the measured spectral window.

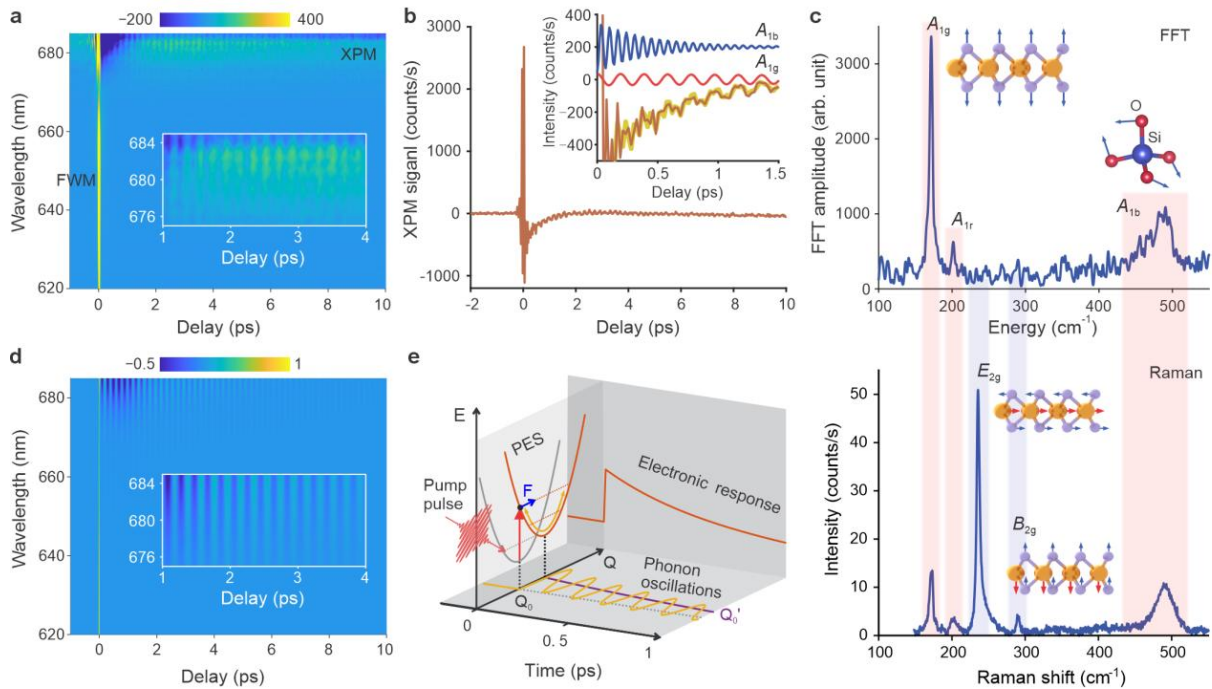

**Figure 2 | Ultrafast modulation of Kerr nonlinearity via coherent phonons. a**, A series of spectra measured as a function of the delay between the pump and the probe pulses. All the spectra were subtracted by the spectrum recorded at -1 ps delay, in order to remove the contribution of the SPM of the probe pulse, to generate a purely XPM spectral trace. SPM of the pump pulse was not measured as it lies relatively far from the edge of the measured spectral window. Inset shows the close up of the measurement after the zero time delay. **b**, Temporal cross-cut of the XPM spectra at ~ 680 nm from the measurement shown in **a**. Pump and probe laser powers were set to be ~ 2.1 mW (26 pJ) and ~ 3.0 mW (37.5 pJ), respectively. Inset: simulation (yellow) of the experimental time trace (orange) by summing up two damped oscillations, representing the $A_{1g}$ (red) and $A_{1b}$ (blue) phonon modes. **c**, Top panel: Fast Fourier transform (FFT) of the time trace in **b** in the delay range of 0.1 ps to 10 ps. The FFT spectrum reveals a strong peak at ~172 cm$^{-1}$, corresponding to the $A_{1g}$ phonon mode of 2$H$-MoTe$_2$, and a broad peak at ~480 cm$^{-1}$, attributed to the $A_{1b}$ phonon mode of the quartz substrate. Bottom panel: Raman spectrum of the 2$H$-MoTe$_2$ sample on quartz, measured with a 633 nm CW laser excitation source. A laser power of 0.89 mW with an acquisition time of 20 s was used in the measurement. **d**, Simulated pump-probe spectra incorporating both XPM and FWM contributions. Inset shows the close-up of the simulation after the zero delay. **e**, Schematic illustration of the displacive excitation of the coherent phonons. Upon excitation with the pump pulse (red), the potential energy surface (PES) of the system shifts within the Born–Oppenheimer approximation, resulting into a new equilibrium position ($Q_0$') of the nuclei. This sudden shift in the PES exerts a fully symmetrical force, $F$, on the lattice, initiating coherent phonon oscillations (yellow). The PES (orange) and equilibrium position of the nuclei (purple) evolve in time.

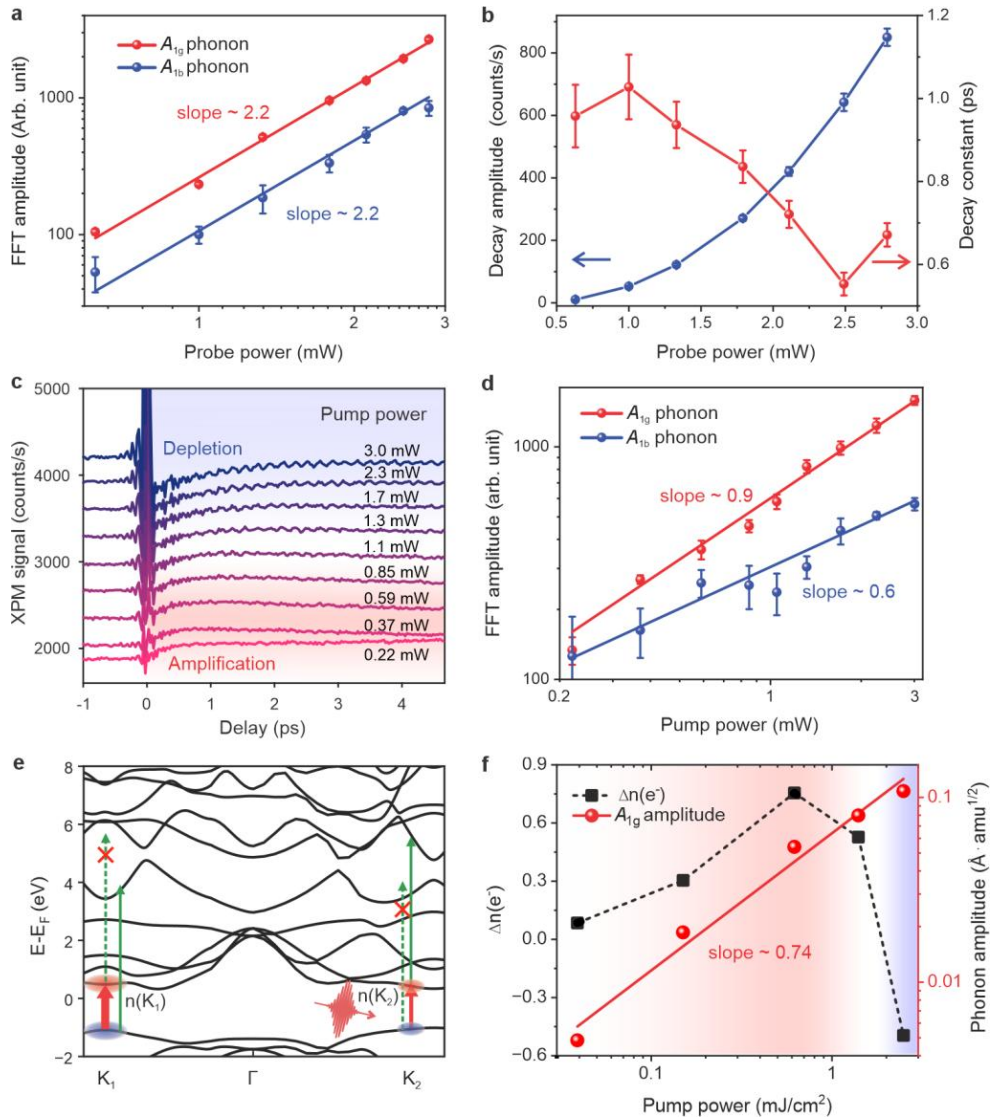

**Figure 3 | Evolution of Kerr nonlinearity with the fluence of the driving pulses and displacive excitation of coherent phonons. a**, Variation of the spectral intensities of the $A_{1g}$ (red dots) and $A_{1b}$ (blue dots) phonon modes on increasing the fluence of the probe pulse. Phonon amplitudes were obtained from the FFT analysis of the time-resolved pump-probe traces (see Fig. S8 in Supplementary Information). Red and blue curves represent fits in the dual logarithmic plots, with fit parameters annotated in the respective legends. The power of the pump pulse was kept at 2.2 mW, and an acquisition time of 0.5 s per delay step was used for all the spectra. **b**, Variation of the amplitudes of the XPM signal at the zero delay between the pump and probe pulses (blue dots), and the decay time constants (red dots) on increasing the fluence of the probe pulse, extracted from the time traces shown in Fig. S8. **c**, A series of time-resolved XPM measurements with increasing fluence of the pump pulse, the XPM signal was enhanced (red area) at low pump fluence but becomes depleted (blue area) at high pump fluence. The fluence of the probe pulse was fixed at 2.1 mW. Pump fluences are annotated to each time trace, which are vertically shifted for clarity. **d**, Variation of the amplitudes of the $A_{1g}$ (red dots) and $A_{1b}$ (blue dots) phonon modes as a function of the increasing fluence of the pump pulse; retrieved from the FFT analysis of the measurements shown in **c**. Red and blue curves display the nonlinear fits in the dual logarithmic plots, with the fit parameters displayed in

the respective legends. **e**, Density functional theory (DFT)-calculated band structure of *2H*-MoTe$_2$. Under pump pulse excitation, electrons at the $K_1$ and $K_2$ points are resonantly promoted to the conduction band (red arrows). The contribution of the probe pulse on XPM arises primarily from resonant interband transitions (solid green arrows). **f**, TDDFT calculated evolution of the population difference of the photoexcited electrons at the $K_1$ and $K_2$ points, $\Delta n = n_{K2} - n_{K1}$ (black square, left y-axis), as a function of increasing fluence of the pump pulses. The evaluated $A_{1g}$ phonon amplitudes (red dots, right y-axis) as a function of the increasing fluence of the pump pulses. The double-logarithmic fit (solid red-curve) of the evaluated phonon amplitudes yields a slope of ~ 0.74, indicating an approximately linear dependence on the pump fluence.

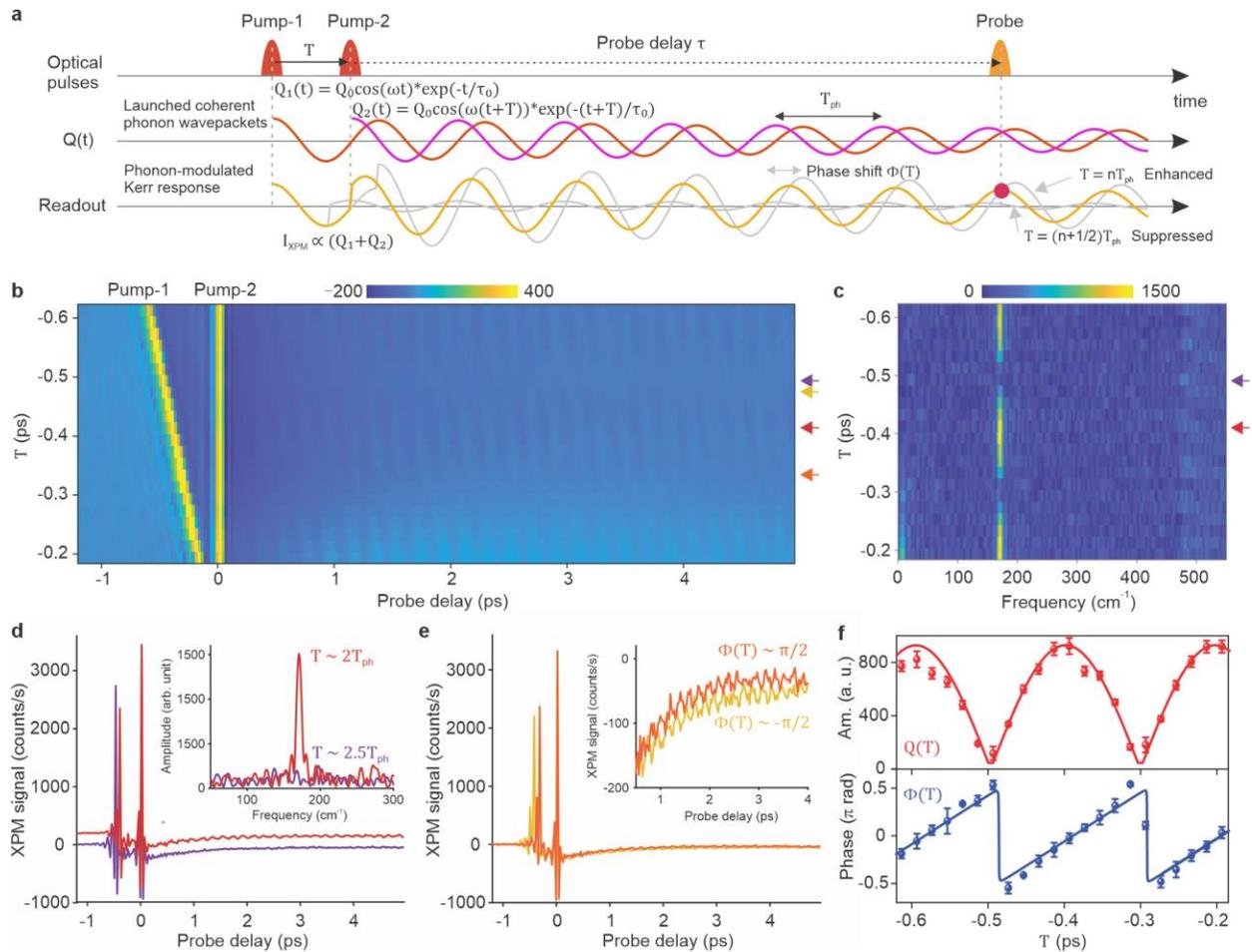

**Figure 4 | All-optical coherent control of Kerr nonlinearity via a dual-pump pulse scheme. a**, Schematic illustration of coherent control phonon dynamics using two pump pulses ('pump-1' and 'pump-2'), which launch coherent phonon wavepackets $Q_1(t)$ (red) and $Q_2(t)$ (magenta), respectively. The relative delay T between two pump pulses controls the phase ($\Phi(T)$) and amplitude (Q(T)) of the launched phonons and, in turn, modulates the phonon-driven Kerr nonlinearity of 2$H$-MoTe$_2$ probed at a later delay $\tau$ by the probe pulse. When $T = nT_{ph}$, the phonon response is enhanced; when $T = (n+\frac{1}{2})T_{ph}$, it is suppressed, where $T_{ph}$ is the phonon oscillation period. **b**, A series of time-resolved traces measured as a function of the delay between the pump-1 (and pump-2) and probe pulses for various delays between the pump-1 and pump-2 pulses (T), as annotated along the y-axis. The temporal crosscuts were extracted at ~ 680 nm from the dual-pump probe measurements. The pump-1 and pump-2 pulse powers were set at 3.4 mW (43 pJ), and the probe pulse power was set at 2.0 mW (25 pJ). The acquisition time was 0.5 s per delay step for all the spectra. **c**, FFT spectra of the time-resolved traces in **b**. **d**, Comparison of the time-resolved traces measured at two fixed delays between pump-1 and pump-2 pulses (T at -0.42 ps and -0.49 ps) as indicated by the red and purple arrows in **b**, demonstrating the intensity control of the $A_{1g}$ phonon mode by varying the delay between the two pump pulses. Inset: FFT spectra of the two time-resolved traces show phonon oscillations being completely activated ($T \sim 2T_{ph}$) or suppressed ($T \sim 2.5T_{ph}$). **e**, Comparison of the time-resolved traces measured at two fixed delays between the pump-1 and pump-2 (T

at -0.31 ps and -0.47 ps) pulses as indicated by the orange and yellow arrows in **b**, demonstrating phase control of the $A_{1g}$ phonon mode by adjusting the delay between the two pump pulses. Inset: A close-up of the phonon oscillations, highlighting the out-of-phase oscillation behavior. **f**, The amplitude (Q(T), upper panel, red dots) and phase ($\Phi$(T), lower panel, blue dots) of the $A_{1g}$ phonon mode in 2$H$-MoTe$_2$ as retrieved from the measurement shown in **b** as a function of T. The solid red and blue curves represent simulations based on two damped harmonic oscillators driven by a pair of pump pulses.